\documentstyle[pre,aps,psfig,twoside,floats,fancyheadings]{revtex}

\pssilent

\begin{document}

\title{Hurst's Rescaled Range Statistical Analysis for Pseudorandom Number\\
       Generators used in Physical Simulations}

\author{B. M. Gammel}

\address{Physik Department der Technischen Universit\"at M\"unchen, 
        T30, 85747 Garching, Germany
\centerline{\rm\small August 7, 1997}\\
\medskip
\parbox{5.515in}{  
\rm\small
\hspace*{1ex}
The rescaled range statistical analysis ($RS$) 
is proposed as a new method to detect
correlations in pseudorandom number generators used in Monte Carlo simulations.
In an extensive test it is demonstrated that the RS analysis provides a very 
sensitive method to reveal hidden long run and short run correlations. 
Several widely used and also some recently proposed pseudorandom number 
generators are subjected to this test. 
In many generators correlations are detected and quantified.
\\[1em]
PACS numbers: 02.70.Lq, 05.40.+j, 02.50.-r, 75.40.Mg \hfill
}
}

\draft
\twocolumn
\maketitle

\section{introduction}
\psfig{figure=preprint.eps,rwidth=0mm,rheight=0mm}\vspace*{-1\baselineskip}
Random numbers are the essential ingredient of all stochastic simulations.
A great many algorithms in Monte-Carlo (MC) simulations and other 
non-physical computational fields rely crucially on the statistical  
properties of the random numbers used.
High precision calculations on nowadays computer hardware typically
involve the generation of billions of random numbers.

Today the most convenient and most reliable method of obtaining random numbers
in practice is the use of a deterministic algorithm. 
Such a numerical method produces
a sequence of {\em pseudorandom numbers} (PRN) which mimic 
the statistical properties of true random numbers as good as possible. 
Usually the pseudorandom number
generator (PRNG) is {\em assumed} to generate a sequence of independent 
and identically distributed continuous ${\rm U}(0,1)$ random number, 
that means uniformly distributed over the interval $(0,1)$.
Other distributions can be obtained by transformation methods
\cite{netrand}.
Since the state space of the generator is finite the sequence of PRNs
will be eventually periodic. Therefore the expected properties of ``true''
random variables can only be approximated.

{\em True random} numbers can only be produced by physical devices that 
generate events which are principally unpredictable in advance, such as noise 
diodes or gamma ray counters. But such devices are inconvenient 
to use and Marsaglia reported that several commercial products 
fail standard statistical tests spectacularly
\cite{marsaglia96,lecuyer97}.
An alternative could be the archiving of random 
numbers of high quality on a CDROM \cite{marsaglia96}, 
although such a source is by far not as convenient 
to handle as a simple function call.

While {\em theoretical test} methods 
\cite{knuth81,lecuyer94}, such as the analysis of
the lattice structure \cite{marsaglia68} of linear congruential generators,
are certainly the starting point for constructing a good PRNG
there is also a strong need for so-called {\em empirical tests}.
These view the PRNG under consideration as a black box 
and statistically analyze sequences of numbers 
for various types of correlations, regardless of the generation method.
There is a large battery of {\em standard tests}
\cite{lecuyer97,knuth81,lecuyer94,james90,niederreiter92,marsaglia96}
which every candidate to be used in ``serious'' simulations has to pass.
PRNGs that have succeeded in all of these tests seemed to work reliable in
apparently all physical simulations until the last few years.
But the rapid development of computer hardware and 
improved simulation algorithms have caused the demands 
on the quality of the random number sequences to greatly 
increase.
As a consequence erroneous results have been found in recent high precision 
MC calculations. The errors could be related to the use of popular 
PRNGs in combination with some specialized algorithms 
\cite{ferrenberg92,selke93,grassberger93,coddington94,schmid95}
which revealed hitherto undetected correlations in 
the pseudorandom sequences.

Thus there is a strong need to enlarge the tool box of empirical tests
to gain confidence in newly proposed PRNGs 
\cite{marsaglia94,lecuyer96a,lecuyer96b,matsumoto94}
and to check whether traditionally used PRNGs are still
reliable in modern applications.
Any good statistical test should have an 
idiosyncracy for unwanted correlations
and detect defects before they show up in an application.
Newly developed and highly specialized algorithms may be sensitive
to structural defects in PRNGs which are not evident in
the standard tests. 
As different tests detect different types of defects 
it is desirable to develop application specific tests 
\cite{vattulainen94,vattulainen95a,vattulainen95b,vattulainen95c}
that are especially sensitive to the features of the random numbers
which are probed in simulations in current fields of research. 
But often this cannot be assessed in advance
and the only way to reassure oneself of the correctness of
a suspicious (or very important) result is to 
perform an {\em in situ test}
and to repeat the simulation with some different PRNGs.
Enlarging the set of test methods
therefore can help to save precious time 
and to avoid painful recalculations.

In section \ref{sec2} a new test method is proposed
which is applied to a set of several popular generators 
described in section \ref{sec3}.
In section \ref{sec4} the results of the numerical experiments are discussed
illustrating the capability of the new test.
Following the conclusions, section \ref{sec5}, additional results are
tabulated in the appendices.

\section{The RS Analysis}\label{sec2}

In the following I describe a new technique for judging the quality of PRNGs 
in at least several physically relevant situations. 
It will be demonstrated that the rescaled  
range statistical analysis ($RS$ analysis) 
provides an extremely  sensitive method for revealing hidden 
correlations in PRNGs.

As this method is based on general
statistical properties expected for an independent Gaussian process
it should also be useful as a general tool to test the suitability of a PRNG
in a wide class of stochastic simulations.
In the sequel it will be shown that it is especially effective
for testing the presence of {\em long run} statistical dependence
and in cases where such a correlation is present, for estimating
its intensity.
In addition it is shown that also {\em short run} cyclic components 
in a pseudorandom sequence are easily made evident using 
the $RS$ statistic.

Hydrology is the oldest discipline in which 
noncyclic long run dependence has been reported.
In particular the $RS$ analysis 
has been invented by Hurst \cite{hurst51,hurst65}
when he was studying the Nile
in order to describe the long term dependence 
of the water level in rivers and reservoirs. 
Later his method has gained much attention in the context 
of fractional Brownian motion \cite{mandelbrot68}.

The $RS$ statistic for a series $\xi_t$ in the discrete integer valued time
is conventionally defined as follows:
\begin{eqnarray}
\label{eq_rs}
X(t,s)  &=& \displaystyle\sum_{u = 1}^t 
              ( \xi_u - \langle\xi\rangle_s )\\ \nonumber
R(s)    &=& \displaystyle\max_{1 \le t \le s} X(t,s) 
                        - \min_{1 \le t \le s} X(t,s) \\ \nonumber 
S(s)    &=& \displaystyle\biggl[ {1\over s} \sum_{t = 1}^s 
                     (\xi_t - \langle\xi\rangle_s )^2 
                                \biggr]^{1\over2}\\ 
RS(s)  &=& \displaystyle R(s) / S(s) \nonumber
\end{eqnarray}

Viewing the $\xi_t$ as spatial increments 
in a one-dimensional
random walk then 
$\sum_{t = 1}^s \xi_t$ 
is the distance of the walker 
from the starting point at time $s$.
In the quantity $X(t,s)$ the mean 
\begin{equation}
\langle\xi\rangle_s = {1\over s} \sum_{t = 1}^s \xi_t 
\end{equation}
over the time lag $s-1$
is subtracted to remove a trend
if the expectation value of $\xi_t$ is not zero.
In the sequel the difference between the final time $s$ and 
the initial time $1$ of the stochastic process 
will be termed the lag $\tau = s-1$.
$R(\tau)$ is the {\em self-adjusted range of the cumulative sums}
and $RS(\tau)$ is the {\em self-rescaled self-adjusted range}, 
which is the quantity of our interest. 

Feller \cite{feller51} has proved
that the asymptotic behavior for the expectation value
of {\em any independent random} process 
with {\em finite variance} is given by
\begin{eqnarray}
  \lim_{\tau\to\infty} {\rm E}[\tau^{-{1\over2}}\,\, RS(\tau)] &=& \sqrt{\pi/2}.
\label{eq_asymptotic}
\end{eqnarray}
The combination $R(\tau)/S(\tau)$ has a better sampling stability
than $R(\tau)$,
in the sense that the relative deviation of $RS$, defined as
$\Delta RS(\tau) = \sqrt{{\rm Var}[RS(\tau)]} \,\,/\,\, {\rm E}[RS(\tau)]$,
is smaller \cite{mandelbrot69}. 
For an independent Gaussian process 
the limiting standard deviation is
\begin{eqnarray}
\lim_{\tau\to\infty}{\rm Var}[RS(\tau)] = \sqrt{\pi^2/6-\pi/2} \approx 0.2723.
\label{eq_asympvariance}
\end{eqnarray}

On the other hand Hurst had found empirically that 
many time series of natural phenomena are described
by the scaling relation 
\begin{eqnarray}
  RS(\tau) &\propto& \tau^H.
\end{eqnarray}
where $H$ differs significantly from $1/2$. 
In the context of 
fractional Brownian motion 
\cite{mandelbrot68,mandelbrot69}
a Hurst exponent of 
$H=1/2$ corresponds 
to the vanishing of correlations between past and 
future spatial increments in the record. 
For $H > 1/2$ one has persistent behavior, that means a 
positive increment for some time  in the past 
will on the average lead to a positive
increment in the future (if the increments are 
distributed symmetrically around zero). 
Correspondingly the case of $H < 1/2$ denotes 
antipersistent behavior.

Thus almost all {\em long run} 
correlations in the stochastic process should show up
in deviations from the asymptotic 
(\ref{eq_asymptotic}), (\ref{eq_asympvariance}).

Furthermore, Mandelbrot and Wallis have demonstrated that 
the value of the asymptotic prefactor $\sqrt{\pi/2}$ is not robust
with respect to {\em short run} statistical dependence \cite{mandelbrot69}.
This value can be arbitrarily modified by cyclic components
in the random process. The superposition of a white noise 
(with zero mean and unit variance) and a
purely periodic process, for instance, leads to
an asymptotic value of $\sqrt{\tau\pi/2}\,\,(1+A/2)^{-1/2}$,
with $A$ being the amplitude of a sine wave. 
Moreover, the transient to the asymptotic is not smooth, 
but typically shows a series of oscillations,
resembling the case of a purely oscillatory process
\cite{mandelbrot69}.

Therefore the $RS$ statistic is perfectly suited to analyze
a stochastic process for correlations on {\em all} scales.

In the following section several types of PRNGs will be used to
generate $U(0,1)$-distributed random numbers $\xi_t$.
The sequence of $\xi_t$ will then be analyzed 
according to the $RS$ statistic.
It will be demonstrated that various PRNGs produce
sequences of numbers that show deviations from the asymptotic
behavior (\ref{eq_asymptotic}), (\ref{eq_asympvariance}).
Moreover, it is found that for finite lags $\tau$ 
the value of $RS(\tau)$ differs significantly 
between the tested PRNGs being indicative of
short range correlations.
This way it is possible to obtain a complete
``fingerprint'' of correlations of a PRNG
and to measure their intensity as a function of the lag.

\section{Random Generators}\label{sec3}

Because of the vast number of different PRNGs currently employed in
simulations only a small fraction can be selected in this work.

The generators of the first group, labeled G1 to G7, are included as
they are in general use -- either because of traditions, 
because they are recommended in popular books, or
because they can be found in many commercial software packages.
Some of them have documented defects (G1,G2,G3,G5). 
These are considered here to study how their deviations show up in the 
$RS$ statistics. The generators in the second group, G8 to G11,
have been proposed recently to match also future requirements on
period length and quality. But there is little documented experience about
their behavior in physical simulations.
As there are many good reviews and books on the various generation
methods and the performance in the standard tests 
\cite{lecuyer97,knuth81,lecuyer94,james90,niederreiter92,%
bratley87,lecuyer90,tezuka95}
only a brief outline of the considered algorithms 
is given in the next section.

\subsection{Generation Methods}

Most of the commonly used PRNGs are based on the linear congruential method.
In general a {\em multiple recursive generator} of order $k$, 
denoted by MRG($a_1,\dots,a_k;c;m$), 
is based on the $k$th-order linear recurrence
\begin{eqnarray}
x_n &=& (a_1 x_{n-1} + \cdots + a_k x_{n-k} + c) \mathop{\rm mod} m,
\end{eqnarray}
where the order  $k$ and the modulus $m$ are positive integers 
and the coefficients are
integers in the range $\lbrace -(m-1),\dots,m-1\rbrace$.
The numbers $x_n$ of the sequence are then scaled to the interval $(0,1)$
by $u_n=x_n/m$. 

The special case, where $k=1$, is
the well-known {\em linear congruential} generator LCG($a;c;m$)
introduced by Lehmer \cite{lehmer51},
or in the homogeneous case, $c=0$, 
the {\em multiplicative linear congruential} generator,
denoted by MLCG($a;m$). It can be shown that 
a recursion of order $k$ with a
non-zero constant $c$ is equivalent to some 
homogeneous recurrence of order $k+1$ 
\cite{lecuyer94,lecuyer90}.
All congruential generators show a pronounced lattice structure. 
That means, if $n$ subsequent numbers are used to form vectors in 
the $n$-dimensional space all points that can be generated within the period
lie on a family of equidistant parallel hyperplanes
\cite{marsaglia68}. 
Tables with good choices for
the constants can be found in recent reviews 
\cite{lecuyer97,lecuyer90,park88,lecuyer88}.

A {\em lagged Fibonacci} generator, LF($l_1,\dots,l_k; m;\circ$), 
with $k$ lags is obtained for 
$c=0$ and $k$ coefficients $a_i$ being set to unit modulus, 
the others being set to zero,
\begin{eqnarray}
x_n &=& (x_{n-l_1} \circ \cdots \circ x_{n-l_k}) \mathop{\rm mod} m.
\end{eqnarray}
The binary operator $\circ$ is usually either addition or subtraction.

The {\em Linear feedback shift register} 
or {\em Tausworthe} method, LFSR($p,q$),
generates a sequence of binary digits (bits) $b_n$ from the recurrence
relation 
\begin{eqnarray}
b_n &=& b_{n-p} \oplus b_{n-q}
\end{eqnarray}
where the exclusive-or operation $\oplus$ is equivalent
to a bitwise addition modulo two 
\cite{niederreiter92,tausworthe65}.
A sequence of pseudorandom numbers is then obtained by 
taking an appropriate number of consecutive bits to form an integer number.

{\em Generalized feedback shift register} generators
\cite{lewis73}, denoted by 
GFSR($\l_1,\dots,l_k;m$),
which can be considered as a generalization of the 
Tausworthe generator,
are related to the lagged Fibonacci
method, but use the exclusive-or operation
instead of the arithmetic operators to combine computer words $w$
\begin{eqnarray}
w_n &=& w_{n-l_1} \oplus \cdots \oplus w_{n-l_k}.
\end{eqnarray}
A generator of this type with two lags (103 and 250) has been made popular by
Kirkpatrick and Stoll and is known as R250 \cite{kirk81,maier91} 
(see also \cite{ferrenberg92}). A particular realization with four lags
has been given by Ziff \cite{ziff92} (for test results see 
\cite{vattulainen94,vattulainen95a,vattulainen95b,vattulainen95c}).
A recently proposed special variant with huge period is the 
{\em twisted\/} GFSR generator, TGFSR \cite{matsumoto94}.

The {\em multiply-with-carry} generator, denoted by 
MWC($a_1,\dots,a_k;c;m$), is defined 
by the recurrence relation
\begin{eqnarray}
x_n &=& (a_1 x_{n-1} + \cdots + a_k x_{n-k} + c_{n-1}) \mathop{\rm mod} m,\\ \nonumber
c_n &=& (a_1 x_{n-1} + \cdots + a_k x_{n-k} + c_{n-1}) \mathop{\rm div} m.
\end{eqnarray}
The {\rm div} denotes an integer division. Here, in contrast to the
MRG a carry (or borrow) $c_n$ is propagated to the next iteration step.

Special cases of the MWC are the 
the {\em add-with-carry}, AWC($l_1,l_2;m$), 
and the {\em subtract-with-borrow}, SWB($l_1,l_2;m$),
generators, which are obtained by setting two coefficients $a_i$ to unit modulus
and all others equal to zero \cite{marsaglia94,marsaglia91}. 
This basically results in a
LF generator with two lags, but with an extra addition 
of a carry 
\begin{eqnarray}
x_n &=& (x_{n-l_1} + x_{n-l_2} + c_{n-1}) \mathop{\rm mod} m,\\ \nonumber
c_n &=& [x_{n-l_1} + x_{n-l_2} + c_{n-1} \ge m].
\end{eqnarray}
In the case of an AWC the bracket indicates the value of 
the carry which is equal to $1$ if the inequality is true, 
and equal to 0 otherwise.
In the case of an SWB the addition operations accordingly 
have to be replaced by subtractions and the borrow is equal to 1 
if the result
of the subtractions becomes negative.
These generators can produce much longer
periods than the underlying LF generators, but have a bad lattice 
structure in dimension $l+1$, ($l$ being the larger of the lags)
\cite{lecuyer97,lecuyer94,couture94}.

The {\em subtraction method}, SUB($c;m$), is based on a simple 
arithmetic sequence
\begin{eqnarray}
x_n &=& (x_{n-1} - c) \mathop{\rm mod} m.
\end{eqnarray}
This method is not suitable by itself, but it may be included
in combination generators 
\cite{james90,marsaglia90}.

The {\em multiplicative quadratic congruential} method, MQC
\cite{knuth81,niederreiter92},
the {\em cryptographic} BBS 
\cite{blum86}
and DES \cite{press92b} generators,
or the {\em inversive congruential generator}, ICG
\cite{eichenauer86}
are only mentioned for completeness, as these have received 
considerable theoretical attention recently.
These new methods have promising features, but the generators
are currently not in common use as there 
is little practical experience with them.

In general the PRNGs with several lags require an initial set of seeds
$x_1,\dots,x_k$ the number of which is determined by the largest lag $k$.
While most generators do not require a special initialization procedure
care has to be taken with the GFSR generators. Here an improper selection of
the seeds can severely affect the quality of the sequence of PRNs
\cite{fushimi89}. Often a congruential generator is used to generate the
initial state.

Tausworthe and LFSR generators which are based on the theory 
of primitive trinomials
form unfavourable structures similar to the lattice structure of LCGs
and have bad statistical properties 
\cite{lecuyer96b,tezuka95}. 
Such simple generators should be avoided
and combined generators should be used instead.

There is strong empirical support that the combination of different
pseudorandom sequences in general leads to an improved statistical behavior
\cite{knuth81,mclaren65}.
The two well-known methods
are the {\em shuffling} of one sequence with another or with itself
\cite{knuth81,niederreiter92}
or the combination by {\em modular addition}
\cite{lecuyer90,lecuyer88}.
Hybrid generators based on the first method are still not
well understood from the theoretical viewpoint
\cite{lecuyer97,lecuyer94}.
The latter method is better understood and is
suited to obtain very long periods.
Adding two sequences modulo the modulus of either of them the period
obtained is the least common multiple of the component periods.
Generators based on such combination methods currently provide 
us with the ``best'' PRNs. Many different kinds of combined generators
have been proposed, see Refs.\
\cite{knuth81,lecuyer94,james90,marsaglia94,%
lecuyer96a,lecuyer96b,lecuyer90,lecuyer88,marsaglia90} 
and references given therein.

Another common method which {\em can} lead to an improvement of a 
generator is a {\em decimation} strategy, that means a number of PRNs is
thrown away before the next random number is delivered. This approach is taken
for instance in the {\tt RANLUX} generator 
\cite{luescher94,james94} which 
significantly improves the defective 
SWB generator {\tt RCARRY} 
\cite{james90,marsaglia91}.
But neither shuffling nor the decimation method may be desirable
if speed considerations are very important 
(see Appendix \ref{appA} for timing results).

In the following the generators subjected to the $RS$ statistical analysis
are described in brief.

\subsection{Tested Generators}

\begin{description}

\item{G1}
is the well-known MLCG($7^5;2^{31}-1$),
which has been proposed as the ''mimimal standard'
against which all other generators should be judged 
\cite{bratley87,park88,schrage79}.
It is also known as 
{\tt GGL} \cite{park88}, 
{\tt CONG} \cite{ferrenberg92},
{\tt ran0} \cite{press92b,press92},
{\tt SURAND} (IBM computers), 
{\tt RNUM} (IMSL library), or
{\tt RAND} (MATLAB software).
It has the serious drawback of a short period, $2^{31}-1$, and a 
pronounced lattice structure in low dimensions. 
Multiplier and modulus are not the optimal choice
considering several figures of merit, see for instance 
\cite{lecuyer97}.
This generator should only be considered as a toy
for experimenting with new test methods
like all other simple congruential and LFSR generators.

\item{G2} is identical to G1, but additionally Bays-Durham 
shuffling in a table of size 32
is used to improve the low-order serial correlations.
Here the implementation {\tt ran1} of Ref.\ 
\cite{press92b,press92} has
been applied. It is included in this test to show 
the influence of shuffling on the $RS$ statistic.

\item{G3}
is a LF($55,24;2^{31};-$) generator which has a period of $2^{55}-1$. 
It has been devised by Mitchell and Moore in 1958 and is described
by Knuth \cite{knuth81} (originally using an add operation). 
This generator (a version of which is implemented in 
\cite{press92b} as {\tt ran3})
is reported to have significant correlations on 
the bit-level and to fail several
physical tests \cite{grassberger93,vattulainen94,vattulainen95a,vattulainen95b,vattulainen95c}.
It is included to demonstrate the effect of short range
correlations on the $RS$ statistic.

\item{G4}
is a modification of the above generator G3. If a decimation
strategy is used, that means, if only every $k$-th number of the sequence 
is used, the generator passes all of the physical tests 
in Ref.\ \cite{vattulainen94,vattulainen95a,vattulainen95b}
(for $k = 2$ and $k = 3$). 
In this work only the case of $k=3$ is considered.

\item{G5}
ist the GFSR(250,103,$2^{32}$) generator R250 proposed by Kirkpatrick and Stoll
\cite{kirk81,maier91}. It has a period of $2^{250}$. While this generator performs
well in the standard statistical tests it is reported to fail several physical
tests \cite{ferrenberg92,vattulainen94,vattulainen95a,vattulainen95b,vattulainen95c}.

\item{G6} The combination generator {\tt RANMAR} proposed by
Marsaglia and Zaman \cite{james90,marsaglia90} has a period of 
about $2^{144}$. It is based on the subtraction modulo $2^{24}$
of a simple arithmetic sequence \\
\hspace*{1em}SUB($7654321;2^{24}-3$)\\
and a subtractive Fibonacci generator\\
\hspace*{1em}LF($97,33;2^{24};-$)\\
The initial state is generated by another 
combination of LCG($53;1;169$) and
a multiplicative three-lag Fibonacci sequence. 
The implementation of 
James \cite{james90} tested here is in wide-spread use and
has been recommended as a ``universal generator''.

\item{G7}
combines the two congruential sequences\\
\hspace*{1em}MLCG($40014;2^{31}-85$) \\ and \\
\hspace*{1em}MLCG($40692;2^{31}-249$)\\
by modular addition
and applies an additional shuffling in a table of 32 entries.
The period is approximately $2^{62}$.
This algorithm has been invented by L'Ecuyer \cite{lecuyer88} and 
implemented by James \cite{james90} (called {\tt RANECU}). 
The additional shuffling has been added 
in the version {\tt ran2} of Press et al. 
\cite{press92b,press92}. 
Many recommendation for the improvement 
(for instance of the speed) of the later version
have been given by Marsaglia and Zaman \cite{marsaglia94}.
They reported that this generator passes all standard tests.
Because of its popularity the implementation 
of Ref.\ \cite{press92b,press92}
has been used in the following $RS$ analysis.

\item{G8} is the recently proposed PRNG {\tt mzran13} 
of Marsaglia and Zaman. It combines \\
\hspace*{1em}LCG($69069,1013904243;2^{32}$)\\ and \\
\hspace*{1em}SWB($2,3;2^{32}-18$)\\
by modular addition
and has a period of about $2^{125}$ \cite{marsaglia94}.
Although the published program takes advantage of the inherent modulo
$2^{32}$ arithmetic of modern CPU's it can easily be made portable 
to CPUs with any larger word size by using bit-masks.

\item{G9}
This is a composite generator of L'Ecuyer \cite{lecuyer96a}
based on the modular addition of the sequences of \\
\hspace*{1em}MRG($0,63308,-183326;0;2^{31}-1$) \\ 
and \\
\hspace*{1em}MRG($86098,0,-539608;0;2^{31}-2000169$).\\
It has a very long period of about $2^{205}$
and a lattice structure with theoretically better properties
than G7 \cite{lecuyer96a}.

\item{G10}
This generator is the maximally equidistributed
three-component Tausworthe generator {\tt taus88}
developed by L'Ecuyer 
\cite{lecuyer96b}
with a period of approximately $2^{88}$.

\item{G11}
The twisted GFSR generator {\tt TT800}
proposed by Matsumoto and Kurita \cite{matsumoto94}
has a huge period of $2^{800}-1$ and is reported to have
excellent equidistribution properties up to a dimension of 25.
This generator is recommended in \cite{lecuyer97}.
The tested version includes Matsumoto's code change of 1996
which improves the lower bit correlations. 

\end{description}

\section{Description of the Test and Results}\label{sec4}

\subsection{The Test Setup}

A few additional words have to be said about the generation 
of the initial seeds for the PRNGs.
As these are (possibly) the only truly random part when generating
pseudorandom numbers some care should be taken.

The following method has been applied, as it corresponds to a common way
random generators are used in practice:

The initial seed is calculated from a combination of some 
obviously truly random events, such as the time and the date 
when the program is started, several system specific (unique)
process identifiers, and the processor clock state.
For this initial seed a sequence of $10^9$ to $10^{10}$ random numbers
is generated and analyzed according to (\ref{eq_rs}). 
Then for some new random seed 
another sequence is obtained and analyzed.

This procedure has been iterated
until the statistical error for the average of $RS(\tau)$ was considered
small enough.
For each of the generators this amounted to $10^{11}$ to $10^{12}$
generated PRNs.

As this approach does not ensure that the generated 
substreams are disjoint it might look safer to {\em split}
the period into disjoint parts. 
This could be done for almost all generators,
but there are several cases known where these (typically) equidistantly 
spaced seeds introduce even worse correlations
\cite{lecuyer94}.
One should also bear in mind that for the long period generators 
there is only a very small probability that, for instance, ten or twenty
sequences of $10^{10}$ numbers selected by a random seed are not disjoint
(of course the period of the ``toy'' generators 
is exhausted immediately).

In the case of generators requiring more than one seed 
{\em one} initial seed has been generated and mixed into the default
seeds of the original source code. 
For instance, the 25 published seeds that define 
the state of the TGFSR generator G11 have all been exclusive-or-ed
with a new random seed every time a new sequence
has been generated.

All calculations necessary to evaluate the $RS$ statistic
have been performed in double precision using IEEE 754 standard
floating point arithmetics.

The number of PRNs generated in the test of each generator
is comparable to the number of random variates
typically required in a nowadays high precision Monte-Carlo simulation.
Such a number may seem large for a mere test, but it comprises
the current state of the art in research fields like percolation,
random walks, diffusion limited aggregation, 
and many others \cite{ferrenberg92,grassberger93,schmid95}.
Considering the speed of the advances in computer technology 
much larger simulations will be in reach within the next few years
posing increased demands on precision to the PRNGs.
Correspondingly the stringency of the empirical tests has to increase too. 

In the following section it will be shown that several current 
thought-to-be-reliable PRNGs show pronounced correlations
in the $RS$ statistic. 
This does not mean that a large scale simulation inevitably produces 
erroneous results with such a PRNG, but it just means that in
some types of simulations deviations are not unlikely 
if high precision is required.
Moreover, the main purpose of this paper is to demonstrate that
the $RS$ statistic is a candidate to enrich the toolbox 
of empirical tests for random number generators.

\begin{figure}[tbh]
\centerline{\psfig{figure=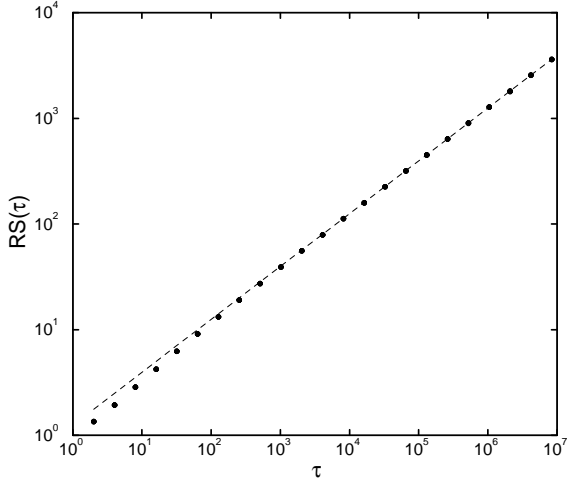,width=75mm}}
\caption{Double-logarithmic plot of the numerical data ($\bullet$) of 
$RS(\tau)$ for all PRNGs. On this scale the results for the various PRNGs are 
indistinguishable.
The asymptotic $\protect\sqrt{\tau\pi/2}$ 
behavior is indicated by the broken line.
}
\label{fig1}
\end{figure}
 
\begin{figure}[tbh]
\centerline{\psfig{figure=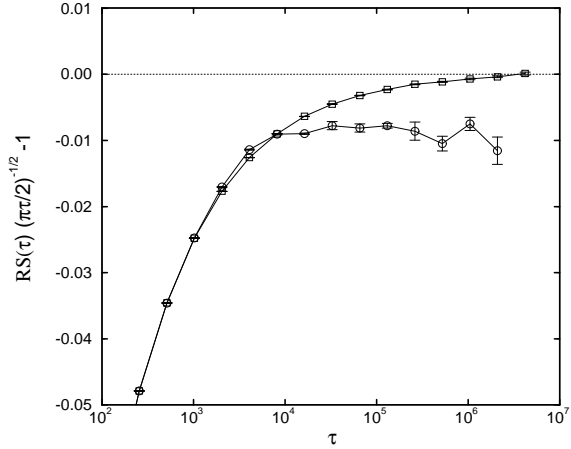,width=75mm}}
\caption{Semi-logarithmic plot of 
$RS(\tau)(\pi\tau/2)^{-1/2} - 1$ for the 
pseudorandom number generators 
G1 ($\circ$) and G9 ($\Box$).
The lines are intended as a guide to the eye.}
\label{fig2}
\end{figure}

\begin{figure}[tbh]
\centerline{\psfig{figure=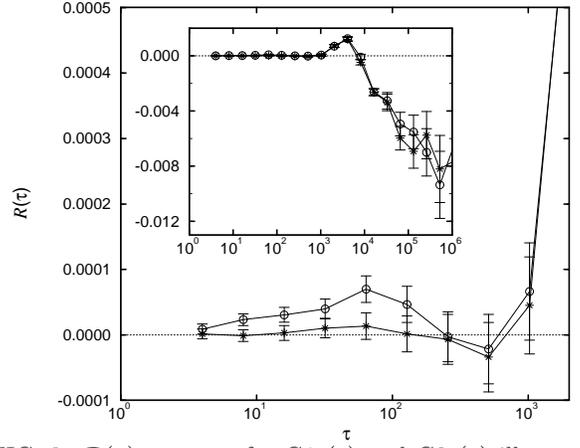,width=75mm}}
\caption{$\protect{\cal R}(\tau)$ versus $\tau$ for
G1 ($\circ$) and G2 ($*$) illustrating the effect of a shuffle table.
The inset shows a larger range of $\tau$.}
\label{fig3}
\end{figure}
 
\begin{figure}[tbh]
\centerline{\psfig{figure=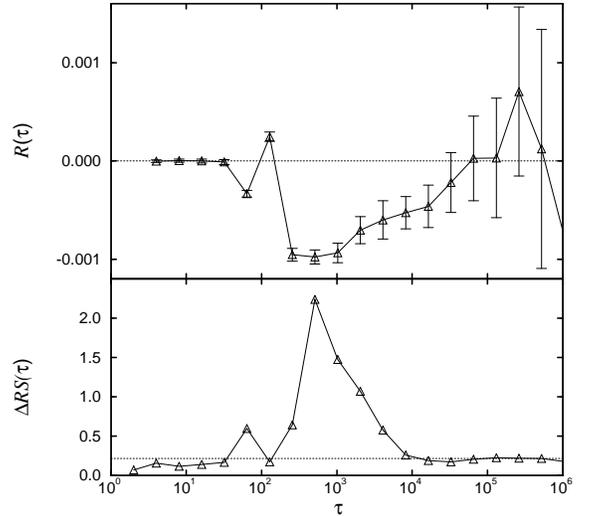,width=75mm}}
\caption{Upper figure: $\protect{\cal R}(\tau)$ versus $\tau$ for the
LF generator G3 ($\bigtriangleup$). 
Lower figure: Drastic deviations from the asymptotic value (dotted line)
are also visible in $\Delta RS(\tau)$.}
\label{fig4}
\end{figure}

\begin{figure}[tbh]
\centerline{\psfig{figure=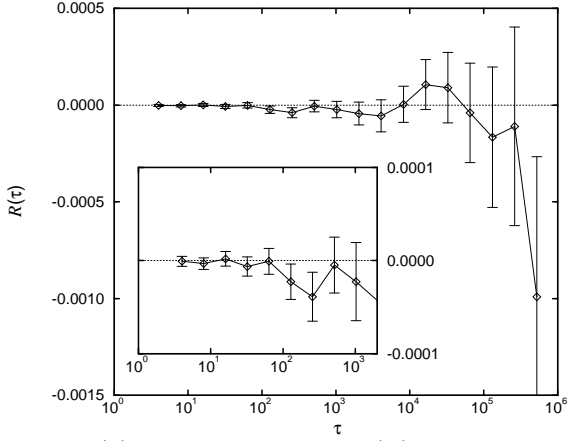,width=75mm}}
\caption{$\protect{\cal R}(\tau)$ for the generator
G4 ($\Diamond$). Inset: Magnified view for small lags $\tau$.}
\label{fig5}
\end{figure}

\begin{figure}[tbh]
\centerline{\psfig{figure=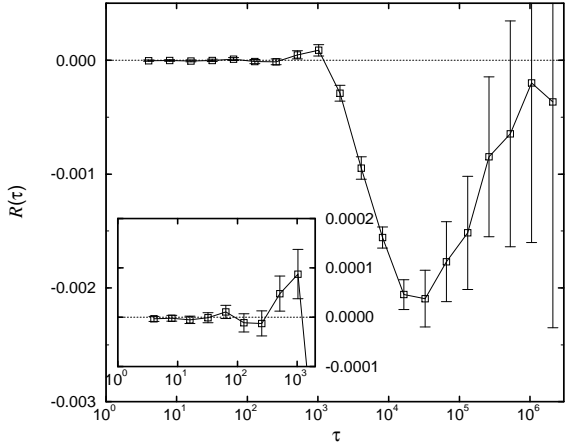,width=75mm}}
\caption{$\protect{\cal R}(\tau)$ for the GFSR generator
G5 ($+$). Inset: Magnified view for small lags $\tau$.}
\label{fig6}
\end{figure}

\begin{figure}[tbh]
\centerline{\psfig{figure=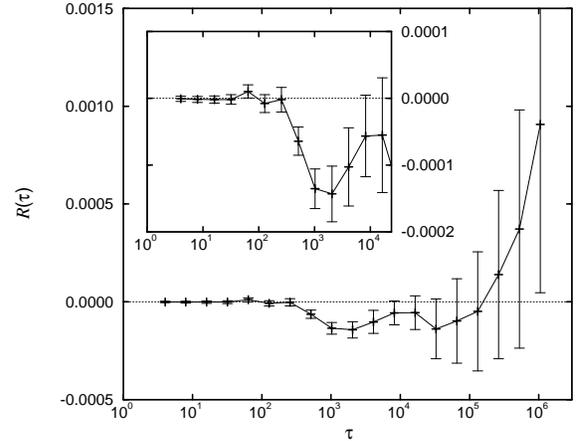,width=75mm}}
\caption{$\protect{\cal R}(\tau)$ for the combination generator
G6 ($\times$). Inset: Magnified view for small lags $\tau$.}
\label{fig7}
\end{figure}

\begin{figure}[tbh]
\centerline{\psfig{figure=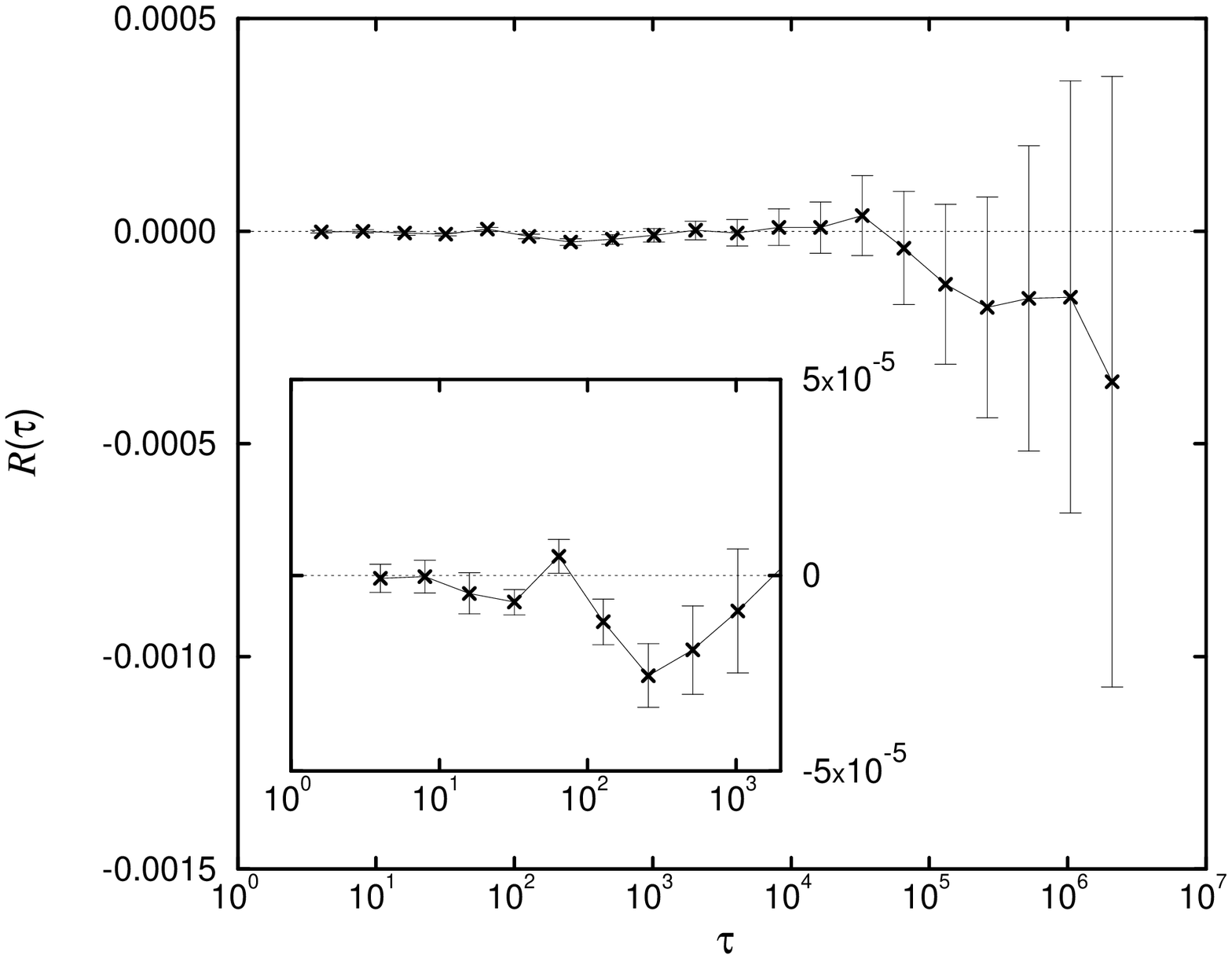,width=75mm}}
\caption{$\protect{\cal R}(\tau)$ for the combination generator
G7 ($\Box$). Inset: Magnified view for small lags $\tau$.}
\label{fig8}
\end{figure}

\begin{figure}[tbh]
\centerline{\psfig{figure=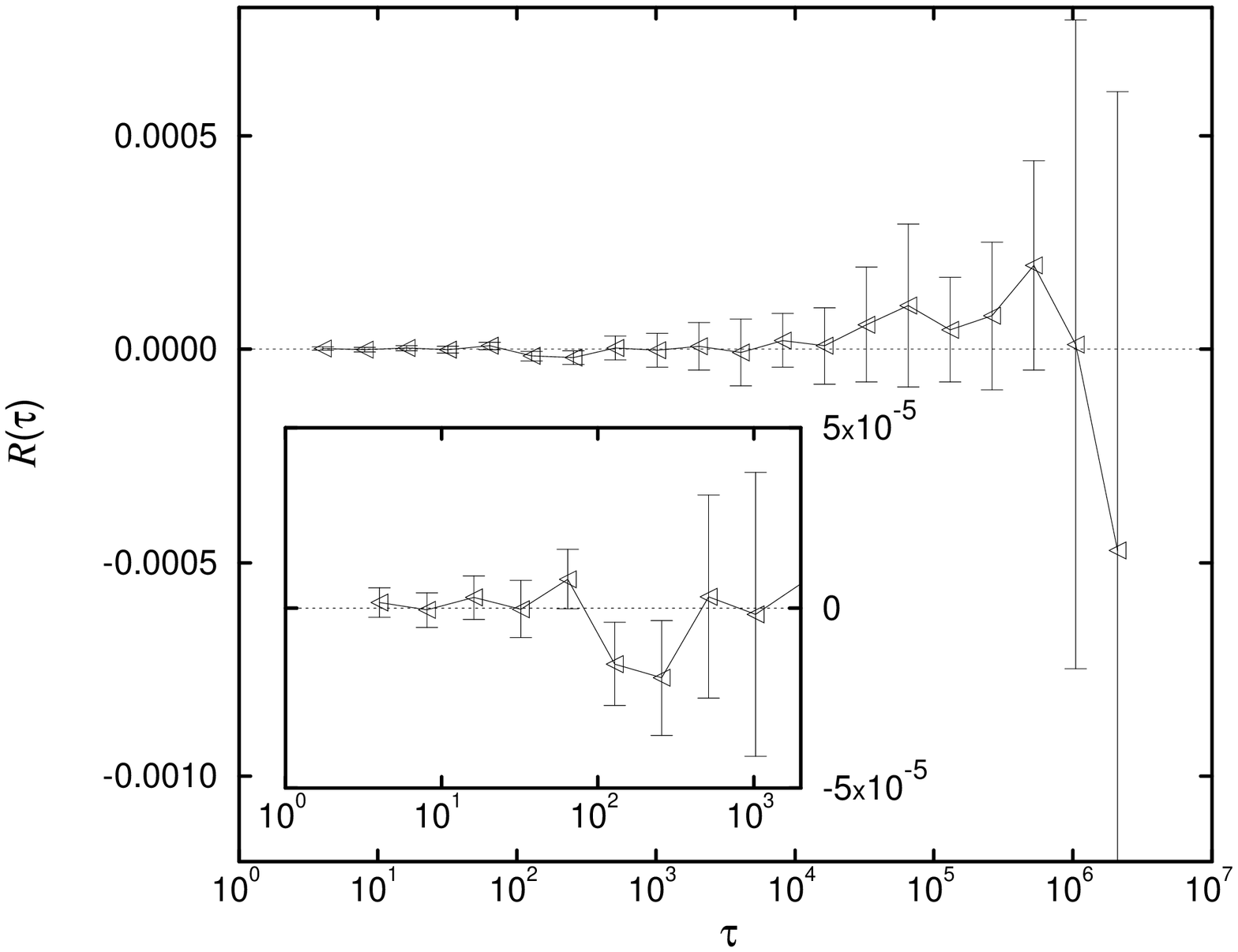,width=75mm}}
\caption{$\protect{\cal R}(\tau)$ for the combination generator
G8 ($\lhd$). Inset: Magnified view for small lags $\tau$.}
\label{fig9}
\end{figure}

\begin{figure}[tbh]
\centerline{\psfig{figure=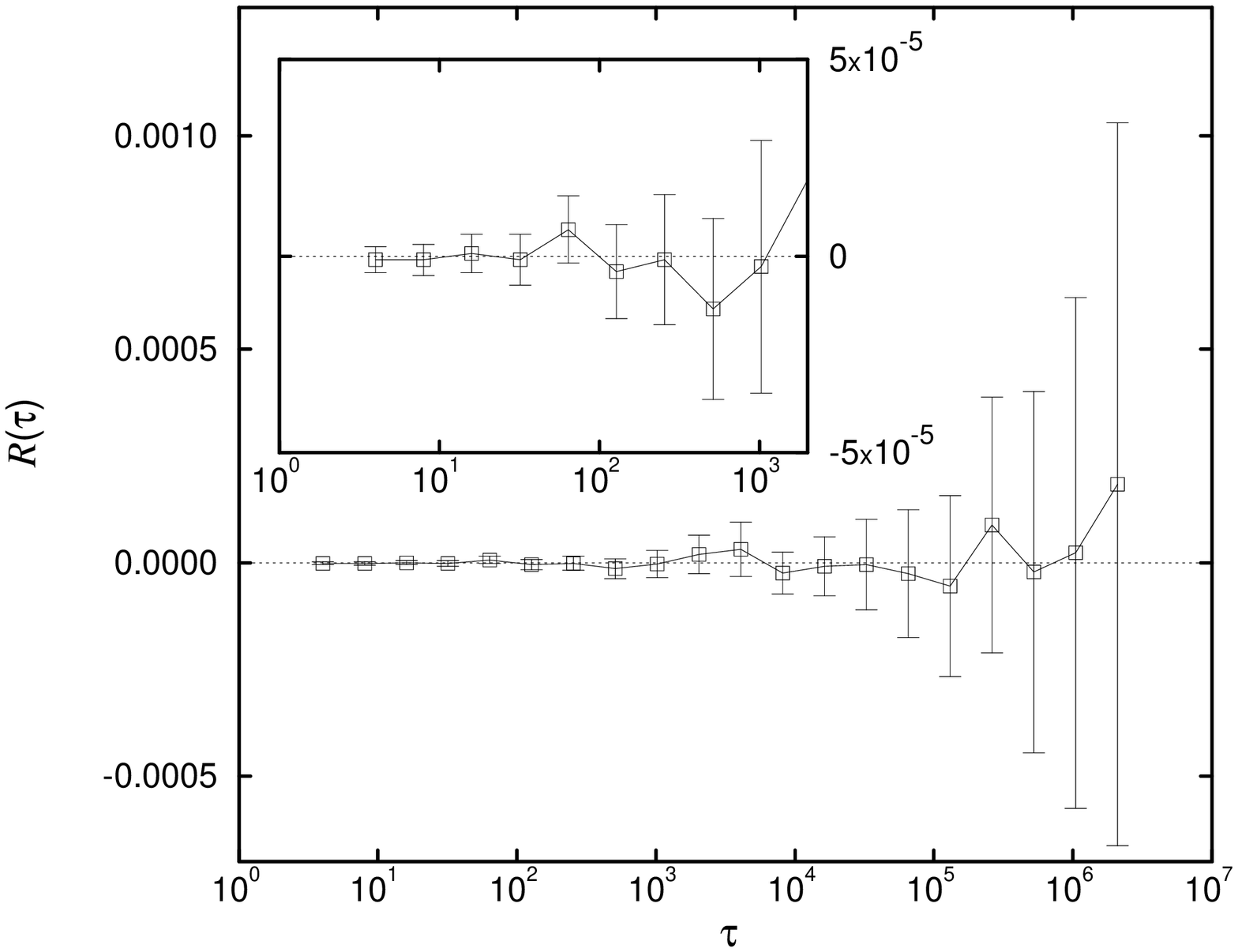,width=75mm}}
\caption{$\protect{\cal R}(\tau)$ for the combination generator
G9 ($\Box$). Inset: Magnified view for small lags $\tau$.}
\label{fig10}
\end{figure}

\begin{figure}[tbh]
\centerline{\psfig{figure=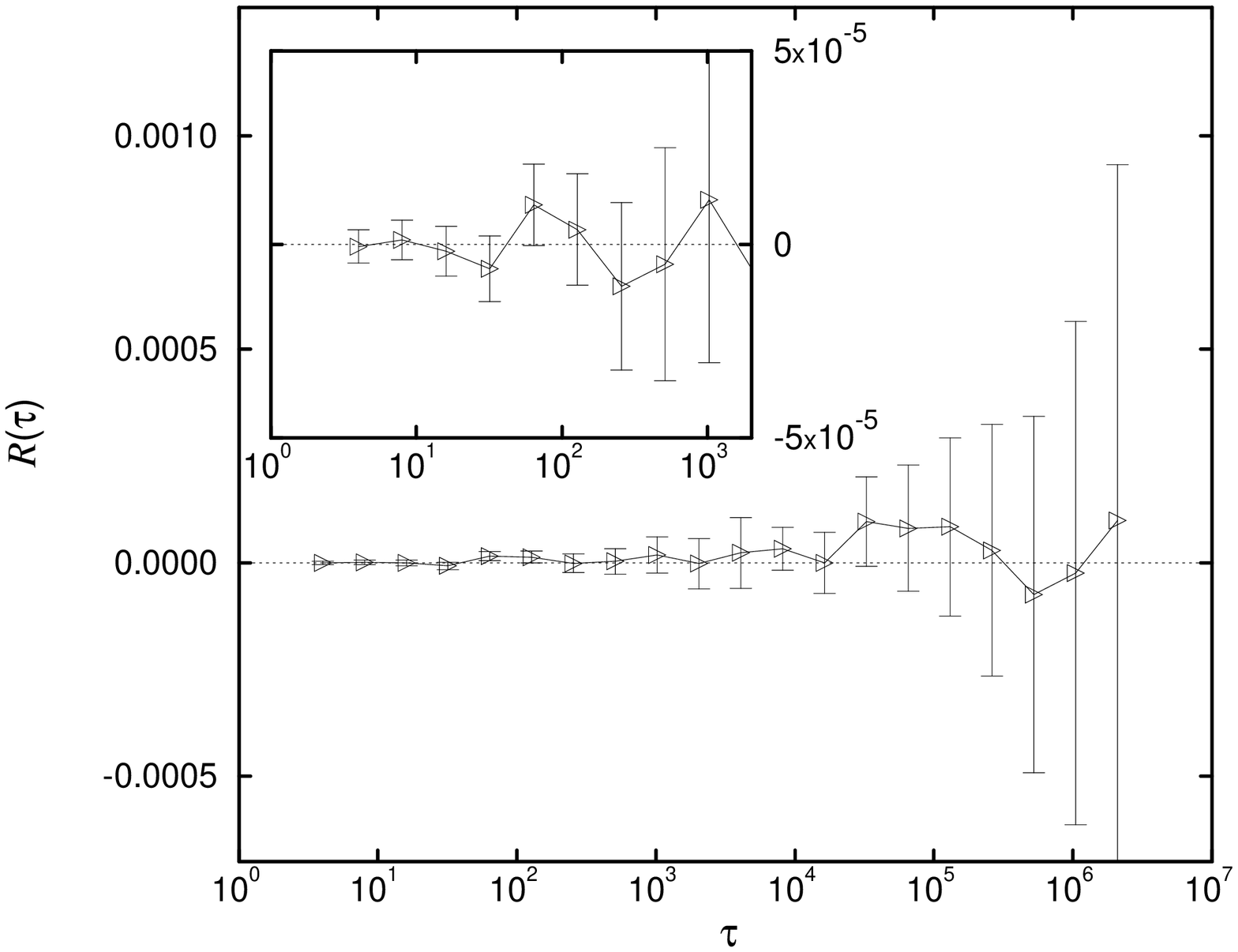,width=75mm}}
\caption{$\protect{\cal R}(\tau)$ for the combination generator
G10 ($\rhd$). Inset: Magnified view for small lags $\tau$.}
\label{fig11}
\end{figure}

\begin{figure}[tbh]
\centerline{\psfig{figure=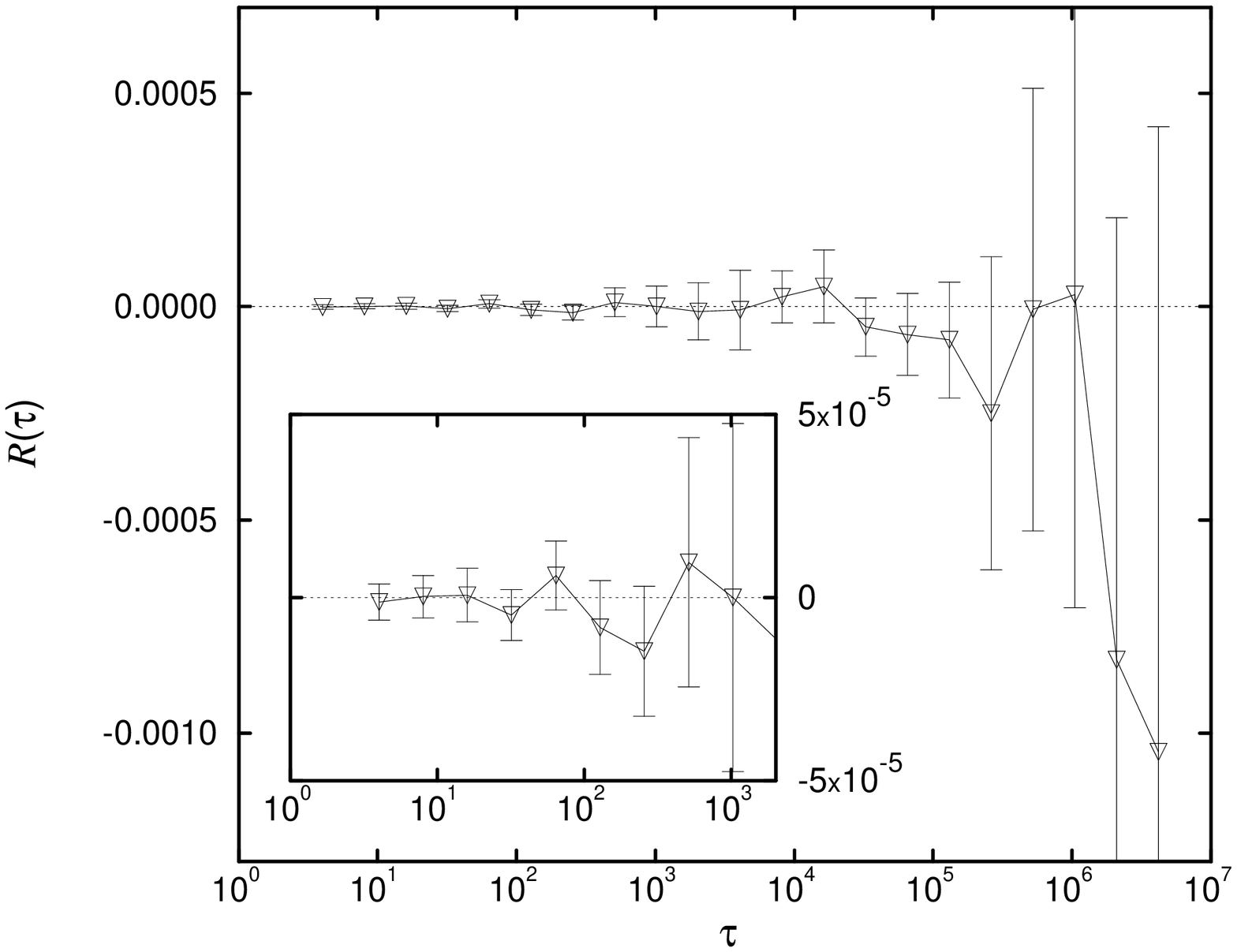,width=75mm}}
\caption{$\protect{\cal R}(\tau)$ for the TGFSR generator
G11 ($\bigtriangledown$). Inset: Magnified view for small lags $\tau$.}
\label{fig12}
\end{figure}

\subsection{Analysis of the RS Data}

In Fig.\ \ref{fig1} the diagram of $\log RS(\tau)$ versus
$\log\tau$ is shown for all tested random generators.  $RS(\tau)$ has been
calculated for all powers of two in the range 
from $\tau=2$ up to $\tau=2^{23} \approx 8\times10^6$ as indicated by the dots.
After a {\em transient} behavior for lags smaller than $\tau \approx 10^4$ the
asymptotic law (\ref{eq_asymptotic}) applies {\em almost} perfectly.
But on this scale the results for the various PRNGs are 
indistinguishable for all lags.

To resolve differences between the PRGNs it is convenient
to remove the asymptotic trend.
In Fig.\ \ref{fig2} the reduced function 
$RS(\tau)(\pi\tau/2)^{-1/2} - 1$ is displayed for a generator with known
correlations, G1 ($\circ$), and the combination generator G9 ($\Box$).
On this  scale of magnification it can be seen that the simple LCG 
spectacularly fails to approach the expected asymptotic.
The relative deviation becomes as large as 1\% corresponding to a reduced
asymptotic prefactor (which appears to be approximately 1.243 instead
of $\sqrt{\pi/2}=1.253$).
For comparison the data for the highly reliable composite MRG G9 are shown.
In this case the asymptotic expectation value is approached smoothly. 
Due to the large statistical ensemble the error bars appear as single lines.

The distribution of the numerical $RS$ values for all lags is
 well described by the 
slightly right-skewed asymptotic density as given by Feller \cite{feller51}.
The half width of the error bars for the estimate of the mean
(in this and the following figures) is given by two standard 
deviations according to the asymptotic analytical result
(\ref{eq_asympvariance}). 
This corresponds to a confidence level of about 95\%.
The numerical results for the mean together with the standard deviation 
of the mean are tabulated in Appendix \ref{appT} for all generators
of this test.

As with several other test statistics where only the asymptotic distribution
is available one is limited to compare the generators. 
Comparing the estimate of the mean for finite lags with the asymptotic
expectation one could always enforce a rejection of a generator if the 
the number of samples is sufficiently increased. 
In the following a method is described which facilitates the comparison
of $RS(\tau)$ for the different generators.

It can be safely assume that the asymptotic limit is approached smoothly
with increasing $\tau$.
Therefore any apparent local and non-monotone structure in the transient 
will be indicative of correlations.
Analyzing the functional form of the transient
a simple and smooth interpolation can be found which gives 
an accurate approximation for all lags within a range
of more than 6 orders of magnitude.
The transient of $RS(\tau)$ can be parametrized by
\begin{eqnarray}\label{eq_fit2}
{\cal R}(\tau) 
 \equiv 
 \biggl(\displaystyle\frac{RS(\tau)}
                 {\sqrt{\displaystyle\pi\tau/2}-\alpha} - 1\biggr)
       &\!\!-\!\!& \biggl(\displaystyle\frac{1}{\arctan{\beta \tau}}
                          -\frac{2}{\pi}\biggr) \\ \nonumber
       &\!\!+\!\!& \gamma e^{\displaystyle -\delta \tau^{\displaystyle\varepsilon}}.
\end{eqnarray}
Using only two parameters $\alpha,\beta$ the first two terms suffice
to approximate the transient with a relative precision of 
$\approx 10^{-5}$ for all lags larger than $\tau = 32$.
The last term in (\ref{eq_fit2}) has been introduced to
approximate the transient for lags as small as $\tau = 4$.
The coefficients have been obtained from a numerical adjustment
using the mean values obtained from the stronger generators 
G8, G9, and G10 with $\tau$ in the range from $4$ to $2^{14}$. 
In this range the individual results agree to a high precision.
The values of the coefficients in (\ref{eq_fit2}) 
used in the following are
\begin{equation}
\begin{tabular}{l@{\quad}l@{\quad}l}
$\alpha\approx 1.0319941$& $\gamma\approx 0.10516938$& 
$\varepsilon\approx 0.61775533$\\
$\beta \approx 0.42091184$& $\delta\approx 0.90187633$
\end{tabular}
\label{eq_data}
\end{equation}
The smooth interpolation ${\cal R}(\tau)$ of the transient now allows 
an unbiased comparison of the various PRNGs.
As the expectation values for finite $\tau$ are not known the approximation
(\ref{eq_fit2}),(\ref{eq_data}) is used instead. The generators can now be
compared with the approximate transient. This approach has been found to be
superior to comparing the generators individually. In particular 
the influence of statistical fluctuations of the mean are minimized
compared to a pairwise comparison of the generators at a given value of $\tau$.
In the following it will become clear that the important point is not
to have a precise approximation of the transient for truly random numbers.
The detection of a deviation is insensitive to the exact form of the
approximation: in all cases a defect showed up as a pronounced 
wiggle in $RS(\tau)$ around the monotone transient.
Therefore the subtraction of any monotone and slowly varying
function would suffice to reveal a characteristic ``fingerprint'' of correlations
in the PRNG.
All systematic deviations of ${\cal R}(\tau)$ from zero 
are indicative of the presence of correlations
and the amplitude at lag $\tau$ can be considered as a measure of the 
strength of correlations for the given lag. Hence the various PRNGs can be
compared quantitatively.

\subsection{Discussion of the Results}

In Fig.\ \ref{fig3} the semi-logarithmic plots of 
${\cal R}(\tau)$ versus $\log\tau$ for the toy generators G1 ($\circ$) 
and G2 ($*$) are shown for lags between 4 and $2^{21} \approx 2\times 10^6$ 
(inset). Serious deviations are evident for lags larger than $10^3$. 
Magnifying on the vertical axis by a factor of 25 the plot of
${\cal R}(\tau)$ reveals deviations also at small lags (main figure).
In generator G2 additional shuffling in a small table 
has been introduced to improve low order serial correlations of generator G1.
For lags up to $\tau \approx 128$ the deviations are indeed strongly reduced. 
As expected there is no improvement for lags which are much 
larger than the size of the shuffling table.

In Fig.\ \ref{fig4} the results  
for the lagged Fibonacci generator G3 ($\bigtriangleup$) are shown.
This generator is known to fail several tests 
(see Ref.\ \cite{vattulainen94,vattulainen95a,vattulainen95b,vattulainen95c} 
and appendix \ref{appC}).
It is reassuring to see that the $RS$ statistic easily reveals the onset of
disastrous correlations at $\tau$ corresponding to the larger lag 
of the generator ($l=55$). 
The deviations show up as a crossover of ${\cal R}(\tau)$ (upper figure)
to a ``shifted asymptotic'' reflecting a modified asymptotic 
prefactor. This gives evidence to the presence of some strong
cyclic components in the pseudorandom process of G3.
This is the only generator in this test showing also deviations of
$\Delta RS(\tau)$ from the asymptotic value (Fig.\ \ref{fig4} lower graph).

If a decimation strategy with $k=3$ is applied,
corresponding to generator G4 ($\Diamond$),
the correlations are strongly suppressed (Fig.\ \ref{fig5}).

The GFSR generator G5 (Fig.\ \ref{fig6}) uses larger lags than G3 shifting the
onset of correlations to larger $\tau$.
The magnitude of the deviation is even twice as large as that of
generator G3. These dramatic deviations are obviously indicators for
the poor behaviour of G5 in some Monte-Carlo (MC) simulations
\cite{vattulainen94}.

Pseudorandom numbers of much better quality are 
expected from combination generators which can overcome the weakness
of generators which are structurally too simple.

In Fig.\ \ref{fig7} the performance of the popular combination
generator G6 ($+$) can be estimated. 
When $\tau$ is somewhat larger than the lags of the LF component of
the generator significant deviations in ${\cal R}$
are observed (similar to G3 and G5). These are presumably due to the
deficient LF component of the composite generator.
But compared to G5 the deviation is about 10 times smaller. 
For the time being there are no documented failures in physical simulations
that use this generator \cite{vattulainen95a}. But comparing
Fig.\ \ref{fig7} with Figs.\ \ref{fig4},\ref{fig6} one can conclude
that deviations in MC simulations
are not implausible if higher precision is demanded.

PRNGs which are as fast, but which have better long-range properties
are discussed in the following.
In the next figure, Fig.\ \ref{fig8}, the results for the 
combined congruential generator G7 ($\times$) are shown.
Compared to the previous generators the amplitude 
of the deviations is drastically decreased.
But for lags in the range $\tau=2^5$ to $2^9$ a structure being indicative of
correlations can be resolved (see inset of figure and Tab.\ \ref{table3})
on a high level of significance.
Although G7 is doubtlessly one of the better generators within this test
it should be immediately evident that it cannot come up to the 
expectations of Press and Teukolsky 
\cite{press92b,press92} 
to provide {\em perfect} random numbers 
(within the limits of its floating point precision).
Thus their proposed ``practical'' definition of {\em perfect}
should at least be put into perspective. 

Random numbers of much better quality (at least in the RS statistics)
are generated by the recently proposed
composite generators G8 to G11. For all lags in the range
$2^2$ to $2^{21}$ there are no significant differences.
These four PRNGs are based on four different generation methods.
Generator G8 applies a combination of generators with different algebraic
structure while the two-component MRG G9 and the three-component Tausworthe
generator G10 combine generators of the same class. 
Finally, G11 is a TGFSR generator which distinguishes itself 
by an extraordinary long period
\cite{note1}. The fact that four generators of completely different algebraic
structure and with theoretically favourable properties give consistent
results reassures that the observed deviations of the other generators
are indicators of real defects.

It should be noted that $RS(\tau)$ necessarily has been
sampled on a coarse grid on the logarithmic scale.
Therefore it is possible that several types of correlations which
would have shown up as a narrow structure 
have not been recognized. Nevertheless the observed 
deviations are intriguing.

\section{Conclusions}\label{sec5}

The sensitivity for correlations on all scales and the
robustness predestinates the $RS$ statistic as a 
tool to catch up defects in pseudorandom number generators.
A practical method has been described which makes it easy to obtain a
characteristic fingerprint of the correlations in a
pseudorandom sequence. The deviations can be described quantitatively 
and the performance of generators for some given range of lags can
be compared.

To illustrate the capability of the $RS$ statistical analysis several
popular generators have been subjected to an extensive test.
The randomness of all tested PRNGs whith known defects
could be refuted. Moreover deviations in several generators which are 
thought to be reliable have been quantified.
Thus the $RS$ analysis has to be considered more stringent 
than many of the previously suggested tests in the sense that 
more generators fail it. 

The selection of a PRNG for a specific simulation
depends on the required level of precision
and on the range of the correlations which may have an impact
on the quantity of interest -- although this often cannot be assessed
in advance. 
But no generator showing a performance inferior to another
generator in several tests should be used any longer if 
it doesn't even distinguish itself at least by speed.
Weak correlations in a current state-of-the-art generator 
(like some of this test) can lead to
erroneous results in a tomorrow high-precision calculation.

\def\P{\hphantom{\mbox{$-$}}} 
\def\N#1{$#1$}                
\def\B#1{\fbox{$#1$}}         
\def\S#1{${}_{#1}$}           
\setlength{\fboxsep}{0pt}     

\begin{table}[thb]
\squeezetable
\begin{tabular}{llll}
$\tau$    &  G1                              & G2                                & G3                             \\
\hline
$2^2$     & \B{\P 8.836(3.84)10^{-6}}\S{2}   &  \N{\P 1.572(3.84)10^{-6}}        & \N{-1.228(6.61)10^{-6}}        \\
$2^3$     & \B{\P 2.331(0.45)10^{-5}}\S{5}   &  \N{-1.166(4.50)10^{-6}}          & \N{\P 3.183(7.76)10^{-6}}      \\
$2^4$     & \B{\P 3.075(0.57)10^{-5}}\S{5}   &  \N{\P 2.972(5.68)10^{-6}}        & \N{\P 1.010(9.79)10^{-6}}      \\
$2^5$     & \B{\P 3.969(0.75)10^{-5}}\S{5}   &  \N{\P 1.045(0.75)10^{-5}}        & \N{-1.026(1.29)10^{-5}}        \\
$2^6$     & \B{\P 6.990(1.01)10^{-5}}\S{6}   &  \N{\P 1.348(1.01)10^{-5}}        & \B{-3.324(0.17)10^{-4}}\S{19}  \\
$2^7$     & \B{\P 4.659(1.38)10^{-5}}\S{3}   &  \N{\P 1.593(13.8)10^{-6}}        & \B{\P 2.483(0.24)10^{-4}}\S{10}\\
$2^8$     & \N{-2.911(19.0)10^{-6}}          &  \N{-6.774(19.0)10^{-6}}          & \B{-9.531(0.33)10^{-4}}\S{29}  \\
$2^9$     & \N{-2.170(2.65)10^{-5}}          &  \N{-3.392(2.65)10^{-5}}          & \B{-9.763(0.35)10^{-4}}\S{27}  \\
$2^{10}$  & \N{\P 6.632(3.71)10^{-5}}        &  \N{\P 4.517(3.71)10^{-5}}        & \B{-9.344(0.49)10^{-4}}\S{18}  \\
$2^{11}$  & \B{\P 7.057(0.52)10^{-4}}\S{13}  &  \B{\P 7.230(0.52)10^{-4}}\S{13}  & \B{-7.032(0.69)10^{-4}}\S{10}  \\
$2^{12}$  & \B{\P 1.249(0.07)10^{-3}}\S{17}  &  \B{\P 1.192(0.07)10^{-3}}\S{16}  & \B{-5.987(0.98)10^{-4}}\S{6}   \\
$2^{13}$  & \N{-8.803(9.78)10^{-5}}          &  \B{-4.659(0.98)10^{-4}}\S{4}     & \B{-5.266(0.82)10^{-4}}\S{6}   \\
$2^{14}$  & \B{-2.627(0.13)10^{-3}}\S{20}    &  \B{-2.637(0.13)10^{-3}}\S{20}    & \B{-4.612(1.08)10^{-4}}\S{4}   \\
$2^{15}$  & \B{-3.263(0.31)10^{-3}}\S{10}    &  \B{-3.378(0.31)10^{-3}}\S{11}    & \N{-2.177(1.53)10^{-4}}        \\
$2^{16}$  & \B{-4.948(0.43)10^{-3}}\S{11}    &  \B{-5.969(0.43)10^{-3}}\S{13}    & \N{\P 2.702(21.5)10^{-5}}      \\
$2^{17}$  & \B{-5.529(0.61)10^{-3}}\S{9}     &  \B{-6.934(0.61)10^{-3}}\S{11}    & \N{\P 3.151(30.4)10^{-5}}      \\
$2^{18}$  & \B{-7.006(0.86)10^{-3}}\S{8}     &  \B{-5.760(0.86)10^{-3}}\S{6}     & \N{\P 7.069(4.30)10^{-4}}      \\
$2^{19}$  & \B{-9.363(1.22)10^{-3}}\S{7}     &  \B{-8.212(1.22)10^{-3}}\S{6}     & \N{\P 1.232(6.08)10^{-4}}      \\
\end{tabular}
\caption{The numerical values of ${\cal R}(\tau)$ are tabulated in columns for the generators G1, G2, G3.
The value of one standard deviation ($\sigma$) of the mean is given in parenthesis. If the deviation is larger than
$2\sigma$ the value is framed and the deviation in units of $\sigma$ is attached to the right.}
\label{table1}
\end{table}

\begin{table}[htb]
\squeezetable
\begin{tabular}{llll}
$\tau$    & G4                             &  G5                              & G6                               \\
\hline
$2^2$     & \N{-8.584(26.4)10^{-7} }       &  \N{-3.322(2.58)10^{-6} }        & \N{-7.225(18.7)10^{-7} }         \\
$2^3$     & \N{-3.268(3.10)10^{-6} }       &  \N{-2.058(3.03)10^{-6} }        & \N{-1.715(2.20)10^{-6} }         \\
$2^4$     & \N{\P 1.845(3.92)10^{-6} }     &  \N{-5.343(3.82)10^{-6} }        & \N{-1.889(2.77)10^{-6} }         \\
$2^5$     & \N{-6.375(5.15)10^{-6} }       &  \N{-1.010(5.03)10^{-6} }        & \N{-1.744(3.64)10^{-6} }         \\
$2^6$     & \N{-7.679(69.4)10^{-7} }       &  \N{\P 1.074(0.68)10^{-5} }      & \B{\P 1.019(0.49)10^{-5} }\S{2}  \\
$2^7$     & \B{-2.283(0.95)10^{-5} }\S{2}  &  \N{-1.135(0.93)10^{-5} }        & \N{-7.898(6.71)10^{-6} }         \\
$2^8$     & \B{-3.893(1.31)10^{-5} }\S{2}  &  \N{-1.254(1.28)10^{-5} }        & \N{-2.120(9.28)10^{-6} }         \\
$2^9$     & \N{-4.791(15.0)10^{-6} }       &  \B{\P 4.778(1.78)10^{-5} }\S{2} & \B{-6.419(1.06)10^{-5} }\S{6}    \\
$2^{10}$  & \N{-2.263(2.10)10^{-5} }       &  \B{\P 8.746(2.49)10^{-5} }\S{3} & \B{-1.354(0.15)10^{-4} }\S{9}    \\
$2^{11}$  & \N{-4.342(2.94)10^{-5} }       &  \B{-2.890(0.35)10^{-4} }\S{8}   & \B{-1.432(0.21)10^{-4} }\S{6}    \\
$2^{12}$  & \N{-5.510(4.14)10^{-5} }       &  \B{-9.464(0.49)10^{-4} }\S{19}  & \B{-1.028(0.29)10^{-4} }\S{3}    \\
$2^{13}$  & \N{\P 4.312(46.7)10^{-6} }     &  \B{-1.557(0.05)10^{-3} }\S{33}  & \N{-5.653(3.05)10^{-5} }         \\
$2^{14}$  & \N{\P 1.065(0.64)10^{-4} }     &  \B{-2.058(0.07)10^{-3} }\S{31}  & \N{-5.522(4.30)10^{-5} }         \\
$2^{15}$  & \N{\P 9.044(9.09)10^{-5} }     &  \B{-2.094(0.12)10^{-3} }\S{16}  & \N{-1.385(0.76)10^{-4} }         \\
$2^{16}$  & \N{-3.961(12.8)10^{-5} }       &  \B{-1.770(0.18)10^{-3} }\S{10}  & \N{-9.816(10.8)10^{-5} }         \\
$2^{17}$  & \N{-1.651(1.81)10^{-4} }       &  \B{-1.516(0.25)10^{-3} }\S{6}   & \N{-4.896(15.2)10^{-5} }         \\
$2^{18}$  & \N{-1.093(2.56)10^{-4} }       &  \B{-8.485(3.51)10^{-4} }\S{2}   & \N{\P 1.387(2.15)10^{-4} }       \\
$2^{19}$  & \B{-9.906(3.62)10^{-4} }\S{2}  &  \N{-6.466(4.96)10^{-4} }        & \N{\P 3.722(3.04)10^{-4} }       \\
$2^{20}$  & \N{-6.947(5.12)10^{-4} }       &  \N{-1.979(7.02)10^{-4} }        & \B{\P 9.075(4.30)10^{-4} }\S{2}  \\
\end{tabular}
\caption{The numerical values of ${\cal R}(\tau)$ are tabulated in columns for the generators G4, G5, G6.
See Tab.\ \protect\ref{table1} for an explanation.}
\label{table2}
\end{table}

\begin{table}[htb]
\squeezetable
\begin{tabular}{llll}
$\tau$    &  G7                             & G8                            & G9                           \\
\hline
$2^2$     & \N{-7.948(17.8)10^{-7} }        & \N{\P 1.458(2.04)10^{-6} }    & \N{-8.946(16.7)10^{-7} }     \\
$2^3$     & \N{-3.115(20.9)10^{-7} }        & \N{-6.022(23.9)10^{-7} }      & \N{-9.815(19.6)10^{-7} }     \\
$2^4$     & \N{-4.627(2.64)10^{-6} }        & \N{\P 2.891(3.02)10^{-6} }    & \N{\P 6.311(24.7)10^{-7} }   \\
$2^5$     & \B{-6.886(1.60)10^{-6} }\S{4}   & \N{-3.553(39.7)10^{-7} }      & \N{-9.777(32.4)10^{-7} }     \\
$2^6$     & \B{\P 4.847(2.15)10^{-6} }\S{2} & \N{\P 7.958(4.12)10^{-6} }    & \N{\P 6.729(4.26)10^{-6} }   \\
$2^7$     & \B{-1.192(0.29)10^{-5} }\S{4}   & \B{-1.554(0.58)10^{-5} }\S{2} & \N{-3.911(5.98)10^{-6} }     \\
$2^8$     & \B{-2.563(0.41)10^{-5} }\S{6}   & \B{-1.939(0.80)10^{-5} }\S{2} & \N{-8.900(82.7)10^{-7} }     \\
$2^9$     & \B{-1.906(0.56)10^{-5} }\S{3}   & \N{\P 3.122(14.1)10^{-6} }    & \N{-1.343(1.15)10^{-5} }     \\
$2^{10}$  & \N{-9.156(7.90)10^{-6} }        & \N{-1.836(19.7)10^{-6} }      & \N{-2.701(16.1)10^{-6} }     \\
$2^{11}$  & \N{\P 2.095(11.1)10^{-6} }      & \N{\P 7.061(27.7)10^{-6} }    & \N{\P 2.019(2.26)10^{-5} }   \\
$2^{12}$  & \N{-3.466(15.6)10^{-6} }        & \N{-7.505(38.9)10^{-6} }      & \N{\P 3.240(3.18)10^{-5} }   \\
$2^{13}$  & \N{\P 9.711(21.5)10^{-6} }      & \N{\P 2.112(3.17)10^{-5} }    & \N{-2.428(2.44)10^{-5} }     \\
$2^{14}$  & \N{\P 8.670(30.1)10^{-6} }      & \N{\P 8.369(44.6)10^{-6} }    & \N{-8.337(34.5)10^{-6} }     \\
$2^{15}$  & \N{\P 3.692(4.71)10^{-5} }      & \N{\P 5.826(6.74)10^{-5} }    & \N{-4.166(53.1)10^{-6} }     \\
$2^{16}$  & \N{-3.956(6.65)10^{-5} }        & \N{\P 1.025(0.95)10^{-4} }    & \N{-2.485(7.50)10^{-5} }     \\
$2^{17}$  & \N{-1.24(0.94)910^{-4} }        & \N{\P 4.591(6.13)10^{-5} }    & \N{-5.423(10.6)10^{-5} }     \\
$2^{18}$  & \N{-1.782(1.30)10^{-4} }        & \N{\P 7.842(8.66)10^{-5} }    & \N{\P 8.842(15.0)10^{-5} }   \\
$2^{19}$  & \N{-1.579(1.79)10^{-4} }        & \N{\P 1.968(1.22)10^{-4} }    & \N{-2.139(21.2)10^{-5} }    \\
$2^{20}$  & \N{-1.544(2.54)10^{-4} }        & \N{\P 1.180(38.0)10^{-5} }    & \N{\P 2.337(29.9)10^{-5} }  \\
$2^{21}$  & \N{-3.535(3.59)10^{-4} }        & \N{-4.704(5.37)10^{-4} }      & \N{\P 1.841(4.23)10^{-4} }   \\
\end{tabular}
\caption{The numerical values of ${\cal R}(\tau)$ are tabulated in columns for the generators G7, G8, G9.
See Tab.\ \protect\ref{table1} for an explanation.}
\label{table3}
\end{table}

\begin{table}[htb]
\squeezetable
\begin{tabular}{llll}
$\tau$   & G10                          & G11                            & \\
\hline
$2^2$    & \N{-5.345(21.8)10^{-7} }     & \N{-1.221(2.46)10^{-6} }       & \\
$2^3$    & \N{\P 1.153(2.56)10^{-6} }   & \N{\P 2.644(28.9)10^{-7} }     & \\
$2^4$    & \N{-1.787(3.23)10^{-6} }     & \N{\P 6.721(36.5)10^{-7} }     & \\
$2^5$    & \N{-6.273(4.25)10^{-6} }     & \N{-4.801(3.49)10^{-6} }       & \\
$2^6$    & \N{\P 1.024(0.53)10^{-5} }   & \N{\P 6.012(4.70)10^{-6} }     & \\
$2^7$    & \N{\P 3.864(7.22)10^{-6} }   & \N{-8.174(6.43)10^{-6} }       & \\
$2^8$    & \N{-1.085(1.08)10^{-5} }     & \N{-1.465(0.89)10^{-5} }       & \\
$2^9$    & \N{-5.065(15.1)10^{-6} }     & \N{\P 9.626(17.0)10^{-6} }     & \\
$2^{10}$ & \N{\P 1.159(2.11)10^{-5} }   & \N{\P 5.613(2377)10^{-8} }     & \\
$2^{11}$ & \N{-6.933(29.6)10^{-6} }     & \N{-1.168(3.34)10^{-5} }       & \\
$2^{12}$ & \N{\P 1.959(4.16)10^{-5} }   & \N{-8.471(46.9)10^{-6} }       & \\
$2^{13}$ & \N{\P 3.068(2.52)10^{-5} }   & \N{\P 2.292(3.04)10^{-5} }     & \\
$2^{14}$ & \N{-1.824(35.6)10^{-6} }     & \N{\P 4.688(4.29)10^{-5} }     & \\
$2^{15}$ & \N{\P 9.589(5.23)10^{-5} }   & \N{-4.816(3.40)10^{-5} }       & \\
$2^{16}$ & \N{\P 7.998(7.39)10^{-5} }   & \N{-6.608(4.80)10^{-5} }       & \\
$2^{17}$ & \N{\P 8.373(10.4)10^{-5} }   & \N{-7.858(6.78)10^{-5} }       & \\
$2^{18}$ & \N{\P 2.910(14.8)10^{-5} }   & \N{-2.508(1.83)10^{-4} }       & \\
$2^{19}$ & \N{-7.407(20.9)10^{-5} }     & \N{-7.072(259)10^{-6} }        & \\
$2^{20}$ & \N{-2.363(29.5)10^{-5} }     & \N{\P 2.755(36.7)10^{-5} }     & \\
$2^{21}$ & \N{\P 9.895(41.7)10^{-5} }   & \N{-8.283(5.18)10^{-4} }       & \\
\end{tabular}
\caption{The numerical values of ${\cal R}(\tau)$ are tabulated in columns for the generators G10, G11.
See Tab.\ \protect\ref{table1} for an explanation.}
\label{table4}
\end{table}

\appendix

\section{Numerical results}\label{appT}

The numerical results for the mean of ${\cal R}(\tau)$, as depicted in previous
figures, are reported in tables \ref{table1} to \ref{table4}. 
The value of one standard deviation of the mean is given in parenthesis.
Values which differ from zero by more than two standard deviations are 
framed and the deviation in units of standard deviations is printed behind the box.

\section{Timing results}\label{appA}

In table \ref{table5} the typical execution times relative to the generator
G1 are given. All generators have been configured 
to deliver one PRN per function call 
and no function code has been inlined. 
Although the figures may scatter between different 
architectures, compilers and optimization options they should be
indicative for the relative performance on work station type computers. 
It should be mentioned that in the case of combined MLCGs and combined MRGs
(G7,G9) a floating point implementation is often much faster than an integer
implementation on many modern CPUs. These versions can compete with the fastest
generators of table \ref{table5}. \cite{lecuyer97b}

\begin{table}[htb]
\begin{tabular}{lr@{\hspace{3em}}|lr}
PRNG & rel.\ time & PRNG & rel.\ time \\
\hline
G1 & $\equiv $ 1   & G7  & $\approx$  2.2 \\
G2 & $\approx$ 1.1 & G8  & $\approx$  0.7 \\
G3 & $\approx$ 0.6 & G9  & $\approx$  2.4 \\
G4 & $\approx$ 1.4 & G10 & $\approx$  0.7 \\
G5 & $\approx$ 0.6 & G11 & $\approx$  0.9 \\
G6 & $\approx$ 1.3 & \\
\end{tabular}
\caption{Relative execution times of the generators considered in this test.}
\label{table5}
\end{table}

\section{Additional results}\label{appC}
 
For comparison the performance of the generators G1--G11
in the recently proposed 
{\em n-block test} and the {\em random walk test} 
\cite{vattulainen94,vattulainen95a,vattulainen95b}
has been calculated.
For the group of PRNGs which have already been considered
in Ref.\ \cite{vattulainen94,vattulainen95a,vattulainen95b} 
the results were reproduced. 
The figures for all generators tested newly are reported 
in Tab.\ \ref{table6}.
According to Ref.\ \cite{vattulainen94,vattulainen95a,vattulainen95b} 
the limit of acceptance 
in the  $\chi^2$-test has been chosen $\chi^2 < 7.815$ 
in the case of the {\em random walk test} 
and $\chi^2 < 3.841$ for the {\em n-block test}.  
A generator is assumed to pass the test 
if in at least two of three independent runs the
value of $\chi^2$ is below the given limit.

The only PRNGs which shows significant deviations from the expected 
distributions are generators G3 and G5. If the decimation strategy 
is used then G3 also passes these tests (corresponds to G4).

These results have to be contrasted with the performance of the PRNGs
in the $RS$ statistical analysis which is much more stringent
in the sense that more generators fail it. 

From the presented figures it is obvious that the walk length (block size)
in these tests is too small (by orders of magnitude) to catch the severe
defects at lags that correspond to the large walk lengths 
in realistic simulations.
It is also evident that it is not sufficient to consider only
a fixed lag as the amplitude of the deviations can vary
strongly with the lag. 
Finally the $RS$ statistic appears to be superior considering its
sensitivity for correlations.

\setlength{\fboxsep}{2pt}     
\begin{table}[htb]
\begin{tabular}{lrrr@{\quad}|rrr}
PRNG  & \multicolumn{3}{c}{{\em $\chi^2$ in random walk test}} 
      & \multicolumn{3}{c}{{\em $\chi^2$ in n-block test}} \\
\hline
G1  & 1.386 &  1.539  &  2.499  & 0.197 &  0.067 &  0.079 \\ 
G2  & 2.131 &  2.889  &  5.127  & 0.009 &  0.026 &  0.014 \\ 
G3  & \B{36.567} & \B{61.235} & \B{44.200}                   
         & 1.607 & 2.161 & 1.104                          \\                  
G4  & 1.402 &  2.225  &  7.080  & 0.982 &  0.801 &  1.002 \\ 
G5  & \B{433.98} & \B{490.93} & \B{424.04}                   
         & \B{515.46} & \B{557.06} & \B{491.57}           \\
G6  & 1.883 &  1.958  &  0.780  & 1.797 &  0.152 &  0.214 \\ 
G7  & 1.764 &  0.329  &  1.093  & 0.397 &  0.488 &  0.002 \\ 
G8  & 2.378 &  1.289  &  3.497  & 0.160 &  0.764 &  0.024 \\ 
G9  & 2.275 &  4.663  &  8.249  & 1.592 &  0.008 &  2.598 \\ 
G10 & 2.634 &  1.699  &  0.979  & 0.325 &  2.550 &  0.341 \\ 
G11 & 2.368 &  3.858  &  0.239  & 0.452 &  0.035 &  0.817 \\ 
\end{tabular}
\caption{
  Results for three runs of the {\em random walk test} 
  (walk length $N=750$ using $10^6$ samples) 
  and of the
  {\em n-block test}
  (block size $N=500$ using $3\times10^6$ samples) 
  \protect\cite{vattulainen94,vattulainen95a,vattulainen95b}. 
  The framed figures indicate a failure in this test.
  }
\label{table6}
\end{table}

\medskip
\centerline{\bf ACKNOWLEDGMENTS}
\medskip\medskip

I would like to thank Pierre L'Ecuyer for many valuable discussions
and a critical reading of the manuscript. Stimulating talks
with Eckhard Pehlke and Ferdinand Evers are also acknowledged. 




\begin{thebibliography}{XXXX99}

\bibitem{netrand}
Software packages for this purpose
can be found for instance at {\tt NETLIB} at
{\tt http://netlib.att.com/ netlib/random/}.

\bibitem{marsaglia96}
G.\ Marsaglia,
in {\em Computer Science and Statistics: The Interface}, Ed.: L.\ Billard,
Elsevier Science Publ., Amsterdam, p.\ 3, 1985;
The software package 
{\tt DIEHARD}, {\em A battery of tests of randomness}, 
is vailable via WWW
at {\tt http://stat.fsu.edu/~geo/diehard.html}, 1996;
The {\tt Marsaglia Random Number CDROM} contains
4.8 billion random bits obtained from a combination
of several sources.

\bibitem{lecuyer97}
P.\ L'Ecuyer,
Chap.\ 4 {\em Random Number Generation} 
in {\em Handbook on Simulation}, Ed. Jerry Banks, Wiley, 1997.

\bibitem{knuth81}
D.\ E.\ Knuth, 
{\em The Art of Computer Programming, Volume 2: Seminumerical Algorithms},
Addison-Wesley, Reading, MA, 2nd edition, 1981.

\bibitem{lecuyer94} 
P.\ L'Ecuyer,
Ann.\ Oper.\ Res.\, {\bf53}, 77, (1994).

\bibitem{marsaglia68}           
G.\ Marsaglia,
Proc.\ of the Nat.\ Acad.\ Sci.\ {\bf 61}, 25 (1968).

\bibitem{james90}               
F.\ James,
Comput.\ Phys.\ Commun.\ {\bf 60}, 329 (1990).

\bibitem{niederreiter92}
H.\ Niederreiter, {\em Random Number Generation and Quasi-Monte Carlo Methods},
Vol.\ 63, SIAM, Philadelphia, 1992.

\bibitem{ferrenberg92}
A.\ M.\ Ferrenberg, D.\ P.\ Landau, and Y.\ J.\ Wong,
Phys.\ Rev.\ Lett.\ {\bf 69}, 3382 (1992).

\bibitem{selke93}
W.\ Selke, A.\ L.\ Talapov, and L.\ N.\ Shchur,
Pis'ma Zh.\ Eksp.\ Teo.\ Fiz.\ {\bf 58}, 684 (1993) 
[JETP Lett.\ {\bf 58}, 665 (193)].

\bibitem{grassberger93}
P.\ Grassberger,
J.\ Phys.\ A {\bf 26}, 2796 (1993); 
Phys.\ Lett.\ A {\bf 181}, 43 (1993).

\bibitem{coddington94}
P.\ D.\ Coddington,
Int.\ J.\ Mod.\ Phys.\ C {\bf 5}, 547 (1994).

\bibitem{schmid95}
F.\ Schmid, N.\ B.\ Wilding,
Int.\ J.\ Mod.\ Phys.\ C {\bf 6}, 781 (1995).

\bibitem{marsaglia94}
G.\ Marsaglia and A.\ Zaman, 
Comput.\ Phys.\ {\bf 8}, 117 (1994); 
Apparently the published C source code for the composite generator 
{\tt mzran13} contains a misprint: 
the ``$-$'' in line 8 should be replaced by a ``$=$''.

\bibitem{lecuyer96a}    
P.\ L'Ecuyer,
Oper.\ Res.\ {\bf 44}, 816 (1996).

\bibitem{lecuyer96b}    
P.\ L'Ecuyer,
Math.\ Comp.\ {\bf 65}, 203 (1996).

\bibitem{matsumoto94}   
M.\ Matsumoto and Y.\ Kurita,
ACM Trans.\ Models and Comput.\ Simul.\ {\bf 2}, 179 (1992);
{\em ibid.}, {\bf 4}, 254 (1994); the 1996 version
which improves the lower bit correlations can be obtained from the author.

\bibitem{vattulainen94}
I.\ Vattulainen, T.\ Ala-Nissila, and K.\ Kankaala, 
Phys.\ Rev.\ Lett.\ {\bf 73}, 2513 (1994).

\bibitem{vattulainen95a}
I.\ Vattulainen, T.\ Ala-Nissila, and K.\ Kankaala, 
Phys.\ Rev.\ E {\bf 52}, 3205 (1995).

\bibitem{vattulainen95b}
I.\ Vattulainen and T.\ Ala-Nissila, 
Comput.\ Phys.\ {\bf 9}, 500 (1995).

\bibitem{vattulainen95c}
I.\ Vattulainen, K.\ Kankaala, J.\ Saarinen, and T.\ Ala-Nissila, 
Comput.\ Phys.\ Commun.\ {\bf 86}, 209 (1995).

\bibitem{hurst51}
H.\ E.\ Hurst, 
Trans.\ Am.\ Soc.\ Civ.\ Eng.\ {\bf 116}, 770 (1951).

\bibitem{hurst65}
H.\ E.\ Hurst, R.\ Black, Y.\ M.\ Sinaika, 
{\em Long-Term Storage in Reservoirs:
An experimental Study}, Constable, London, 1965.

\bibitem{mandelbrot68}
B.\ B.\ Mandelbrot and J.\ W.\ van Ness,
SIAM Review {\bf 10}, 422 (1968).

\bibitem{feller51}
W.\ Feller, 
Ann.\ Math.\ Stat.\ {\bf 22}, 427 (1951);

\bibitem{mandelbrot69}
B.\ B.\ Mandelbrot and J.\ R.\ Wallis,
Water Resour.\ Res.\ {\bf 5}, 967 (1969).

\bibitem{bratley87}
P.\ Bratley, B.\ L.\ Fox, and L.\ E.\ Schrage,
{\em A Guide to Simulation}, 2nd edition, Springer-Verlag, New York, 1987.

\bibitem{lecuyer90}    
P.\ L'Ecuyer,
Commun.\  ACM {\bf 33}, 85 (1990).

\bibitem{tezuka95}
S.\ Tezuka,
{\em Uniform Random Numbers: Theory and Practice},
Kluwer Academic Publishers, Norwell, Mass.\, 195.

\bibitem{lehmer51}
D.\ H.\ Lehmer,
in {\em Proc. 2nd Symp.\ on Large-Scale Digital Calculating Machinery},
Harvard Univ.\ Press, Cambridge, p.\ 141 (1951).

\bibitem{park88}
S.\ K.\ Park and K.\ W.\ Miller,
Comm.\ ACM {\bf 31}, 1192 (1988).

\bibitem{lecuyer88}    
P.\ L'Ecuyer,
Commun.\  ACM {\bf 31}, 742 (1988).

\bibitem{tausworthe65}
R.\ C.\ Tausworthe,
Math.\ Comp.\ {\bf 19}, 201 (1965).

\bibitem{lewis73}
T.\ G.\ Lewis, W.\ H.\ Payne,
J.\ Assoc.\ Comput.\ Mach.\ {\bf 20}, 456 (1973).

\bibitem{kirk81}               
S.\ Kirkpatrick and E.\ Stoll,
J.\ Comput.\ Phys.\ {\bf 40}, 517 (1981).

\bibitem{maier91}               
W.\ L.\ Maier, Dr.\ Dobb's Journal, 
May, 152 (1991). 

\bibitem{ziff92}               
R.\ M.\ Ziff,
Phys.\ Rev.\ Lett.\ {\bf 69}, 2670 (1992).

\bibitem{marsaglia91}           
G.\ Marsaglia and A.\ Zaman,
Ann.\ Appl.\ Probability {\bf 1}, 462 (1991).

\bibitem{couture94}
R.\ Couture, P.\ L'Ecuyer,
Math.\ Comp.\ {\bf 62}, 798 (1994).

\bibitem{marsaglia90}           
G.\ Marsaglia, A.\ Zaman, and W.\ W.\ Tsang,
Stat.\ Prob.\ Lett.\ {\bf 8}, 35 (1990).

\bibitem{blum86}
L.\ Blum, M.\ Blum, and M.\ Schub,
SIAM J.\ Comput.\ {\bf 15}, 364 (1986).

\bibitem{press92b}
W.\ H.\ Press, B.\ P.\ Flannery, S.\ A.\ Teukolsky, and W.\ T.\ Vetterling,
{\em Numerical Recipes in C}, 2nd edition, Cambridge University Press, 1992.

\bibitem{eichenauer86}
J.\ Eichenauer and J.\ Lehn,
Statist.\ Hefte {\bf 27}, 315 (1986);
J.\ Eichenauer-Hermann,
Int.\ Stat.\ Rev.\, {\bf 60}, 167 (1992);
{\em ibid.}, {\bf 63}, 247 (1995).

\bibitem{fushimi89}
M.\ Fushimi,
Appl.\ Math.\ Lett.\ {\bf 2}, 135 (1989).

\bibitem{mclaren65}
M.\ D.\ McLaren and G.\ Marsaglia,
J.\ Assoc.\ Comput.\ Mach.\ {\bf 12}, 83 (1965).

\bibitem{luescher94}            
M.\ L\"uscher,
Comput.\ Phys.\ Commun.\ {\bf 79}, 100 (1994).

\bibitem{james94}               
F.\ James,
Comput.\ Phys.\ Commun.\ {\bf 79}, 111 (1994).

\bibitem{schrage79}
L.\ E.\ Schrage,
ACM Trans.\ Math.\ Soft.\ {\bf 5}, 132-138 (1979).

\bibitem{press92}
W.\ H.\ Press and S.\ A.\ Teukolsky, 
Comput.\ Phys.\ {\bf 6}, 522 (1992).

\bibitem{lecuyer97b}
P.\ L'Ecuyer, private communication.

\bibitem{note1}
{\em Note added in proof:\/}
The recently porposed ``Mersenne Twister'' of the authors of G11
(a variant of the TGFSR  with a giant period $2^{19937}-1$) has also been
found to pass the RS-test. But one should keep in mind 
that a long period is a necessary, but not a sufficient condition
for a reliable generator.

\end{thebibliography}
\end{document}